\documentclass[aps,prd,onecolumn,showpacs,amsmath,amssymb]{revtex4}
\usepackage{graphicx}
\usepackage{dcolumn}
\usepackage{bm}

\begin{document}

\newcommand{\Gm}[3]{\, {}_{#1}^{\vphantom{#3}} G_{#2}^{#3}}
\newcommand{\Spy}[3]{\, {}_{#1}^{\vphantom{#3}} Y_{#2}^{#3}}
\newcommand{\kap}[3]{\, {}_{#1}^{\vphantom{#3}} \kappa_{#2}^{#3}}
\newcommand{\gama}[3]{\, {}_{#1}^{\vphantom{#3}} \gamma_{#2}^{#3}}
\newcommand{\alfa}[2]{\, \alpha_{#1}^{#2}}


\title{B polarization of the CMB from Faraday rotation}

\author{Claudia Sc\'occola}
\affiliation{Facultad de Ciencias Astron\'omicas y Geof{\'\i}sicas\\
Universidad Nacional de La Plata\\ 
Paseo del Bosque S/N, 1900 La Plata, Argentina}
\email{cscoccola@fcaglp.unlp.edu.ar}

\author{Diego Harari and Silvia Mollerach}
\affiliation{CONICET, Centro At\'omico Bariloche\\
Av. Bustillo 9500, 8400 Bariloche, Argentina}
\email{harari@cab.cnea.gov.ar, mollerach@cab.cnea.gov.ar}

\date{\today}

\begin{abstract}

We study the effect of Faraday rotation due to a uniform
magnetic field on the polarization of the cosmic microwave
background (CMB). Scalar fluctuations give rise only to  
parity-even E-type polarization of the CMB. However in the presence of 
a magnetic field, a non-vanishing parity-odd B-type polarization 
component is produced through Faraday rotation. 
We derive the exact solution for the E and B modes generated by 
scalar perturbations including the Faraday rotation effect of a uniform
magnetic field, and evaluate their cross-correlations 
with temperature anisotropies. 
We compute the angular autocorrelation function of the B-modes in the 
limit that the Faraday rotation is small. We find that uniform
primordial magnetic 
fields of present strength around $B_0=10^{-9}$G rotate E-modes into B-modes 
with amplitude comparable to those due to the weak gravitational lensing effect 
at frequencies around $\nu=30$ GHz. The strength of B-modes produced by Faraday 
rotation scales as $B_0/\nu^2$. 
We evaluate also the depolarizing effect of Faraday rotation upon the cross 
correlation 
between temperature anisotropy and E-type polarization.

\end{abstract}

\maketitle

\section{Introduction}

The detection of a parity-odd component of the CMB polarization, 
called B-mode, can provide a unique tool to probe sub-dominant sources of
CMB perturbations, which are primarily due to scalar energy-density 
fluctuations. Scalar perturbations do not produce B polarization, which is only 
excited by either tensor or vector modes \cite{se97,ka97}. 
It is thus important to identify all potential sources  
of B polarization. 
A significant background of B-modes comes from the effect of 
gravitational lensing on the CMB by the matter distribution, which
transforms E into B polarization \cite{zalsel98}.
Cosmological gravitational waves, as those predicted by inflationary models of 
the early universe, would also imprint  B-polarization upon the CMB.
It has been pointed out that the signature of an inflationary 
gravitational-wave background can only be detected by B-polarization 
measurements if the tensor to scalar ratio $r \ge
10^{-4}$, which corresponds to an energy scale of inflation 
larger than $3\times 10^{15}$ GeV \cite{knox,kesden}, because otherwise
their effect would be obscured by the weak lensing background.  
Quite recently, however, a better technique to {\it clean}
polarization maps from the lensing effect has been proposed, 
which would allow tensor-to-scalar ratios as low as $10^{-6}$, 
or even smaller, to be probed \cite{hirata,seljak03}. 
Other secondary contributions to the B-type polarization arising
during the reionization stage, though of much smaller amplitude,
have been considered in \cite{hu00}. B-modes of polarization are also
produced by secondary vector and tensor modes generated by the
non-linear evolution of primordial scalar perturbations \cite{mo04}.
Primordial magnetic fields can also generate B-mode polarization upon 
the CMB, due to the vorticity they induce upon the photon-baryon fluid 
\cite{se01,ma02,su03}.

We will consider here another source of B-type polarization, that is
the Faraday rotation in a magnetic field of initial E-type polarization. 
If a primordial magnetic field is present at the time
of the last scattering of the microwave photons, it causes 
Faraday rotation of the direction of the linear polarization \cite{ko96},
which generates B-modes from initial E-modes of polarization \cite{sc97}. 
In Ref. \cite{sc97} a cross-correlation between the parity-odd B-modes 
generated by Faraday rotation and the parity-even temperature anisotropies 
$\Theta$ was estimated under simplifying approximations. 
$\Theta$-B or E-B correlations also exist if there are other parity violating 
processes, for instance 
in the case in which the vorticity in the photon-baryon fluid is induced by 
helical magnetic fields generated by parity-violating electroweak interactions 
\cite{po02,ca03}, or if parity-violating interactions additional to those in the
standard electroweak theory exist \cite{lu99}.

Large scale magnetic fields have been observed in galaxies, clusters
and superclusters, but their origin is still unknown 
\cite{kr94,wi02}. They may have formed from dynamo amplification of a
small initial seed field \cite{dyn}, or alternatively they can arise from
the adiabatic compression during structure formation of a cosmological 
field of order $10^{-10} - 10^{-9}$ G today \cite{prim}. 
The existence of these very large scale cosmological fields, although is not 
excluded by present observations, is difficult to be determined observationally. 

Here we derive the exact solution for the E and B modes generated by
scalar perturbations including the Faraday rotation effect of a uniform
primordial magnetic field, and discuss their properties. 
We evaluate the effect of Faraday rotation upon the cross correlation $\Theta$-E 
and the induced correlation $\Theta$-B. We then calculate, 
to first order in the Faraday rotation effect, the angular correlation function 
of the B-modes induced by Faraday rotation of scalar E-modes of polarization in 
a standard cosmological model, and compare it with that due to other potential 
sources of B-modes, namely weak gravitational lensing, gravitational waves,
and the effect of Faraday rotation due to magnetic fields in clusters 
\cite{ta01,oh03}. 

\section{Faraday rotation of CMB polarization}

Polarization of the CMB arises from Thomson scattering of anisotropic
radiation by free electrons. One way of describing linear polarization
is by means of the Stokes parameters $Q$ and $U$ \cite{chandra}.
A magnetic field induces a rotation of the
polarization plane of a linearly polarized wave as it propagates through
an ionized medium. This effect is known as Faraday rotation. If there were a
primordial magnetic field at the recombination epoch, or after the Universe
was reionized, it would induce Faraday rotation on the CMB photons,
mixing $U$ and $Q$ modes of polarization. The evolution equations for their
Fourier modes read \cite{ko96,ha97,zh95}

\begin{eqnarray}
\dot Q + ik\mu Q &=& -\dot \tau Q +\frac{3}{4}\dot\tau(1-\mu^2)S_p +2\omega_B U 
\nonumber\\
\dot U + ik\mu U &=& -\dot \tau U - 2\omega_B Q \, ,
\label{QU}
\end{eqnarray}
where $\mu = \hat{n} \cdot \hat{k}$ is the cosine of the angle between
the photon direction and the Fourier wave-vector,  
and $\dot \tau$ is the differential 
optical depth ($\dot\tau=n \sigma_T a/a_0$
with $\sigma_T$ the Thomson scattering cross section, $n$  
the free electron density, and the derivative is taken with respect to conformal 
time $d\eta=dt/a(t)$). 
We define $\omega_B = 3 \vec{B} \cdot \hat{n} c^2 \dot\tau/16 \pi^2 \nu^2 e$
as the Faraday rotation rate of radiation with frequency $\nu$ in a magnetic field 
$\vec B$ ($c$ is the speed of light and $e$ the electron charge).
Polarization is sourced by quadrupole anisotropies 
in the temperature fluctuations $\Theta$. $S_p$ in eq. (\ref{QU}) is the source term, 
which reads 
$S_p=(\Theta_2+Q_2)/5+Q_0$ if  $\Theta$ and $Q$ are expanded in terms of Legendre 
polinomials as $\Theta=\sum_\ell (-i)^\ell \Theta_\ell P_\ell(\mu)$ 
(and a similar expansion for $Q$)\footnote{Our convention here differs by a factor 
$(i)^\ell (2\ell +1)$ from that of \cite{ko96,ha97,zh95}, and coincides with that
of \cite{hw}.}.

The equations above have the following formal integral solutions
\begin{equation}
(Q\pm iU)(\vec{k},\eta_0) = \frac{3}{4} (1- \mu^2) \int_0^{\eta_0} d\eta 
e^{-ik\mu(\eta_0-\eta)}\dot\tau e^{-\tau} S_p(\eta) e^{\mp i F\tau \hat{B}
\cdot \hat{n}}\, . 
\label{integral}
\end{equation}

Here $\tau(\eta)=\int_\eta^{\eta_0}{\rm d}\eta^\prime \dot\tau(\eta^\prime)$ 
is the optical depth between time $\eta$ and present time $\eta_0$, and 
the parameter $F$ is defined as 

\begin{equation}
F=\frac{3}{8 \pi ^2} \frac{Bc^2}{\nu ^2 e}\approx 0.7 \left(\frac{B}{10^{-9}{\rm G}}
\right )\left(\frac {10 {\rm GHz}}{\nu}\right)^2\, .
\label{F}
\end{equation} 

$F$ gives a measure of the typical Faraday rotation between collisions.  Notice that the 
strength of a primordial magnetic field is expected to evolve in time $\propto a^{-2}$ 
due to flux conservation. Since $\nu\propto a^{-1}$, then $F$ (expressed here 
numerically in terms of the present values of $B$ and $\nu$) does not change with time.

\section{E and B modes}

The generation of temperature and polarization anisotropies in the CMB from gravitational 
perturbations has been studied in detail by several authors \cite{se97,ka97,hw,hu98}. 
A simple and powerful formalism is the total angular momentum method \cite{hw,hu98}, 
which includes the effect of scalar, vector and tensor modes on an equal footing. 
We are going to extensively use the results and notation of Ref. \cite{hw}. 

The temperature and polarization fluctuations are expanded in normal modes
that take into account the dependence on both the angular direction of 
photon propagation ${\hat n}$ and the spatial position ${\vec x}$ as
\begin{equation}
\begin{array}{rcl}
\Theta(\eta,{\vec x},{\hat n}) &=& \displaystyle{
\int \frac{d^3 k}{(2\pi)^3} }
        \sum_{\ell,m}\Theta_\ell^{(m)} \Gm{0}{\ell}{m} \, , \\
 (Q \pm i U)(\eta,{\vec x},{\hat n}) &=& \displaystyle{\int \frac{d^3k}
 {(2\pi)^3}}
        \sum_{\ell,m} 
        (E_\ell^{(m)} \pm i B_\ell^{(m)}) \, \Gm{\pm 2}{\ell}{m} \,, 
\end{array}
\label{EB}
\end{equation}
with spin $s=0$ describing the temperature fluctuation and $s=\pm 2$
describing the polarization tensor and $m=0, \pm 1, \pm 2$ 
corresponding to scalar, vector and tensor perturbations, respectively.
$E_\ell^{(m)}$ and $B_\ell^{(m)}$ are the angular moments of the
electric and magnetic polarization components and  
\begin{equation}
\Gm{s}{\ell}{m}({\vec x},{\hat n})  =
        (-i)^\ell \sqrt{ \frac{4\pi}{2\ell+1}}
        \Spy{s}{\ell}{m}({\hat n}) e^{i{\vec k} \cdot {\vec x}}\, ,
\label{G}
\end{equation}
is the basis of the expansion, in terms of the spin-weighted spherical
harmonics. 

The set of Boltzmann equations for the evolution of
$\Theta_\ell^{(m)}$, $E_\ell^{(m)}$ and $B_\ell^{(m)}$, as well as
their integral solutions are deduced in Ref. \cite{hw}. 
In the presence of a uniform magnetic field, Faraday rotation modifies
the Boltzmann equations satisfied by $E_\ell^{(m)}$ and
$B_\ell^{(m)}$. We quote in the Appendix the modified equations for
completeness. 
We can nevertheless sidestep them inverting equation (\ref{EB}) to
derive an integral solution for $E_\ell^{(m)} \pm i B_\ell^{(m)}$ from
the integral solution for $Q \pm i U$ in eq. (\ref{integral}) 
\begin{equation}
E_\ell^{(m)} \pm i B_\ell^{(m)} =
(i)^\ell\frac{3}{2}\sqrt{\frac{2\ell+1}{4\pi}}
\int_0^{\eta_0}d\eta \dot\tau e^{-\tau} P^{(0)} 
\int d\Omega_{\hat{n}}\Spy{\pm 2}{\ell}{m \ast}({\hat n}) (1-\mu^2)
e^{-ik\mu (\eta_0-\eta)} e^{\mp i F\tau \hat{B}\cdot \hat{n}}\, ,
\label{EpmiB}
\end{equation}
where the source term $S_p$ has been expressed in terms of 
$P^{(0)}=(\Theta_2^{(0)}-\sqrt{6}E_2^{(0)})/10=S_p/2$  \cite{hw}.
First the exponential with the magnetic field is expanded in terms of 
spherical Bessel functions and ordinary spherical harmonics as
\begin{equation}
e^{\mp i F\tau \hat{B}\cdot \hat{n}}=\sum_{\ell,m}i^{\mp\ell}4\pi 
j_\ell(F\tau)Y_\ell^m(\hat n)Y_\ell^{m\ast}(\hat B)\, .
\label{YB}
\end{equation}
Then the product of one ordinary and one spin-weighted spherical harmonic 
is written as a sum of terms proportional to just one $\Spy{s}{\ell}{m}
({\hat n})$ using the Clebsh-Gordan relations for the addition of angular momentum
\cite{hw}:
\begin{eqnarray}
\left(\Spy{s_1}{\ell_1}{m_1} \right)
\left(\Spy{s_2}{\ell_2}{m_2} \right)&=&
\frac{\sqrt{(2\ell_1 + 1)(2\ell_2 + 1)}}{4\pi} \sum_{\ell,m,s}
\left< \ell_1, \ell_2 ; m_1, m_2 | \ell_1, \ell_2 ; \ell, m \right>
\nonumber\\
&&\times
\left< \ell_1, \ell_2 ; -s_1, -s_2 | \ell_1, \ell_2 ; \ell, -s \right>
\sqrt{\frac{4\pi}{2\ell+1}}\, \left( \Spy{s}{\ell}{m}\right)  \, .\nonumber
\label{CG}
\end{eqnarray}

The term $\exp{(-ik\mu (\eta_0-\eta))}$ is also expanded in terms of spherical Bessel 
functions,
with the $\hat e_3$ axis chosen in the direction of $\vec k$ (which leaves only 
$Y_\ell^0(\hat n)$ in the expansion) and the angular integration is
performed. The final result is

\begin{eqnarray}
E_\ell^{(m)} \pm i B_\ell^{(m)} &=& (-1)^{m+1}\sqrt{24\pi}(2\ell+1)\int_0^{\eta_0}
d\eta \dot\tau e^{-\tau} P^{(0)} \sum_{\ell' \ell''}(i)^{\ell-\ell''\mp \ell'}
Y_{\ell'}^{m\ast}
(\hat{B})j_{\ell'}\! (F\tau)\sqrt{2\ell'+1}\times \nonumber\\
&& \times <\ell,\ell';-m,m|\ell,\ell';\ell'',0><\ell,\ell';\pm2,0|\ell,\ell';\ell'',\pm2> 
\epsilon_{\ell''}^{(0)}(k(\eta_0-\eta))\, .
\label{integralEB}
\end{eqnarray}
where $\epsilon_{\ell}^{(0)}(x)=\sqrt{3(l+2)!/8(l-2)!}j_\ell (x)/x^2$.

Using the relation $<\ell,\ell';-2,0|\ell,\ell';\ell'',-2>=
(-1)^{\ell''-\ell+\ell'}<\ell,\ell';+2,0|\ell,\ell';\ell'',+2>$ between Clebsh-Gordan 
coefficients, individual expressions for $E_\ell^{(m)}$ and $iB_\ell^{(m)}$ immediately 
follow. They are similar to the right hand side of eq. (\ref{integralEB}) choosing the 
upper sign, with an extra factor $(1 + (-1)^{\ell''-\ell})/2$ in the case of 
$E_\ell^{(m)}$ and  $(1 - (-1)^{\ell''-\ell})/2$ in the case of $iB_\ell^{(m)}$
inside the sum.

Notice that in the absence of a magnetic field the standard solution \cite{hw} for 
$E_\ell^{(0)}$ is recovered, and also $B_\ell^{(m)}=0$ as it should. Instead, in the 
limit of relatively large magnetic field ($F>>1$), both $E_\ell^{(m)}$ and $B_\ell^{(m)}$
tend to zero due to the rapidly oscillatory nature of the spherical Bessel functions. 
This is a manifestation of the depolarizing effect of differential 
Faraday rotation \cite{ha97}.

The presence of the magnetic field, which breaks the invariance under
rotations, gives rise to a coupling between different $m$ multipoles.
This is evident from eq. (\ref{integralEB}), where a full range of $m$
polarization multipoles is obtained from the initial scalar $m=0$
modes, as well as from the Boltzmann equations displayed in the 
Appendix. 

\section{Faraday rotation effect upon cross-correlations}

In order to calculate correlation functions one has to integrate over wavevectors 
$\vec k$, for which our expressions for $\Theta_\ell^{(m)}$, $E_\ell^{(m)}$ and 
$B_\ell^{(m)}$, valid in a frame with $\hat e_3\parallel \vec k$, need to be rotated 
to a frame with fixed axis.  In the absence of a magnetic field, statistical 
homogeneity and isotropy makes the angular integrations simple, and the angular power 
spectrum of the correlation functions can be calculated as

\begin{equation}
C_{\ell}^{XZ} = \frac{2}{\pi} \int k^2dk \sum_m
 \frac{X_\ell^{(m)*}(\eta_0,k)Z_{\ell}^{(m)}(\eta_0,k)}
{(2\ell+1)^2} 
\end{equation}
with $X,Z=\Theta,E,B$. Statistical homogeneity and isotropy is reflected in the fact that
\begin{equation}
<X_{\ell m}Z_{\ell' m'}>=\delta_{mm'}\delta_{\ell\ell'}C_\ell^{XZ}\, ,
\label{deltas}
\end{equation} 
where $X_{\ell m}$ are the 
coefficients of the multipole expansions in the fixed basis:
\begin{equation}
\begin{array}{rcl}
\Theta(\hat n) &=& 
        \sum_{\ell,m}\Theta_{\ell m} Y_{\ell m}  \\
 (Q \pm i U)(\hat n) &=& 
        \sum_{\ell,m} 
        (E_{\ell m} \pm i B_{\ell m}) \Spy{\pm 2}{\ell}{m} \,. 
\end{array}
\label{EB2}
\end{equation} 

A uniform magnetic field breaks invariance under rotations and defines a preferred 
direction in space. We can take the direction of the magnetic field as the $\hat e_3$ 
direction of the fixed basis. The multipole coefficients $E_{\ell m} \pm i 
B_{\ell m}$ in this basis can be calculated through a rotation of those given by 
eq.~(\ref{integralEB}), as 
\begin{equation}
E_{\ell m} \pm i B_{\ell m}=i^\ell\frac{4\pi}{2\ell+1}\int\frac{{\rm d}^3\vec k}{2\pi^3}
e^{i\vec k \cdot \vec x}\sum_{m'}(E_\ell^{(m')} \pm i B_\ell^{(m')})\Spy{-m'}{\ell}{m*}(\hat k)\, .
\label{EBrotated}
\end{equation}
Here $E_\ell^{(m')} \pm i B_\ell^{(m')}$ are given by eq.~(\ref{integralEB}) with 
$\hat B$ replaced by $\hat k$ in the argument of $Y_{\ell '}^{m'}$, since now the angle 
between $\vec B$ and $\vec k$ is measured from the fixed $\hat e_3$ axis along $\vec B$. 
The factor summed over $m'$ in eq. (\ref{EBrotated}) corresponds to the
rotation of the polarization multipoles corresponding to each
wavevector $\vec k$ mode to the fixed axis system, multiplied by
$\sqrt{(2\ell+1)/4\pi}$. The other factors arise from the definition
of the basis $\Gm{\pm 2}{\ell}{m}$ used in eq.~(\ref{EB}). 
The product of two spherical harmonics appearing after replacing
eq.~(\ref{integralEB}) in this expression can be expanded with the 
relations for addition of angular momentum, with the help of the
relation $(-1)^m\Spy{m}{\ell}{0}(\theta,\phi)=Y_{\ell}^{m}(\theta,0)$,
and the sum over $m'$ can be done using the orthogonality
properties of the Clebsch-Gordan coefficients. The result can finally be expressed in 
terms of the analogous multipole coefficients in the absence of a magnetic field
as \footnote{If the fluctuations were such that B-modes exist in the absence of a 
magnetic field, then equation (\ref{EBfinal}) is easily generalized with 
$\tilde E_{\ell '' m}^{(\ell ')}\pm i\tilde B_{\ell '' m}^{(\ell ')}$ 
in its right hand side.}
  
\begin{equation}
E_{\ell m} \pm i B_{\ell m}=\sum_{\ell',\ell''}\left[\frac{(-1)^{\ell''-\ell}+1}{2}
\pm \frac{(-1)^{\ell''-\ell}-1}{2}\right ]
R(\ell m,\ell',\ell'' m)\tilde E_{\ell ''m}^{(\ell')}
\, .
\label{EBfinal}
\end{equation}
Here  $\tilde E_{\ell m}^{(\ell')}$ coincides with the expression for
the multipole 
coefficients due to scalar fluctuations in the absence of a magnetic field, except 
for an extra term $(2\ell '+1)j_{\ell '}(F\tau)$ inside the time-integral that
defines it. Explicitly:
\begin{equation}
\tilde E_{\ell m}^{(\ell ')}=-i^\ell\frac{4\pi}{2\ell+1}\int\frac{{\rm d}^3\vec k}{2\pi^3}
e^{i\vec k \vec x}\Spy{}{l}{m*}(\hat k)\sqrt{6}(2\ell +1) 
\int_0^{\eta_0} d\eta \dot\tau e^{-\tau} (2\ell ' +1)j_{\ell'}(F\tau) P^{(0)}(\eta)
\epsilon_\ell^{(0)}(k(\eta_0-\eta))\, .
\label{Etildeprime}
\end{equation}
The coefficients $R(\ell m,\ell',\ell'' m'')$ are given, in terms of Clebsch-Gordan coefficients,
by
\begin{equation}
R(\ell m,\ell',\ell'' m'')=i^{\ell'}\frac{\sqrt{2\ell''+1}}
{\sqrt{2\ell +1}}
<\ell',\ell'';m-m'',m''|\ell'\ell'';\ell,m><\ell',\ell'';0,2|\ell'\ell'';\ell,2>\, .
\label{R}
\end{equation}
  
Correlation functions including the Faraday rotation effect can now be immediately 
calculated in terms of their values in the absence of the magnetic field, as 
given by eq.~(\ref{deltas}).
The multipoles of the cross-correlation between temperature anisotropies $\Theta$ 
and  E and B modes can be written as
\begin{equation}
<\Theta^*_{\ell_1m_1}(E_{\ell_2 m_2} \pm i B_{\ell_2 m_2})>=
\left[\frac{(-1)^{\ell_2-\ell_1}+1}{2}
\pm \frac{(-1)^{\ell_2-\ell_1}-1}{2}\right ]\delta_{m_1 m_2}\sum_{\ell'}
R(\ell_2 m_2,\ell',\ell_1 m_1)C_{\ell_1}^{\Theta\tilde E^{(\ell')}}\, .
\label{cross}
\end{equation}
This expression is strictly valid in a frame with $\hat e_3\parallel
\vec B$, and the fact that only objects with the same $m$ are correlated
reflects the cylindrical symmetry of the problem around this axis. The
cross-correlation in any other system can be obtained by noticing that
the multipoles in a new system are related to the ones in the 
$\hat e_3\parallel \vec B$ system  by 
$X'_{\ell m}=\sqrt{4\pi/2\ell+1}\sum_{m'} 
X_{\ell  m'} \Spy{-m'}{\ell}{'m*}(\hat B)$. 
In the general system the cross correlation results
\begin{equation}
<\Theta'^*_{\ell_1m_1}(E'_{\ell_2 m_2} \pm i B'_{\ell_2 m_2})>=
\left[\frac{(-1)^{\ell_2-\ell_1}+1}{2}
\pm \frac{(-1)^{\ell_2-\ell_1}-1}{2}\right ]\sum_{\ell'}
R(\ell_2 m_2,\ell',\ell_1 m_1) \sqrt{\frac{4\pi}{2\ell'+1}}
\Spy{}{\ell'}{'m_2-m_1*}(\hat B)
C_{\ell_1}^{\Theta\tilde E^{(\ell')}}\, .
\label{crossrot}
\end{equation}
Thus, in a general system the correlations between different $m$
values do not vanish. This fact can be used to identify the
presence of a
symmetry axis, that in this case corresponds to the direction of the
magnetic field. 

It is clear from the expressions in eqs.~(\ref{cross}) and
(\ref{crossrot}) that Faraday rotation in a uniform magnetic
field generates non-diagonal $(\ell_1\neq\ell_2)$ correlations 
$\Theta$-B out from initially diagonal correlations $\Theta$-E. 
The difference
between $\ell_1$ and $\ell_2$ must be odd, and contributions to the case
$|\ell_2-\ell_1| = 2k+1$ are increasingly suppressed for larger
$k$. They are in fact  of order larger or equal than $2k+1$ in the
parameter $F$. 
This is because the coefficient $R$ vanishes if 
$|\ell_2-\ell_1|>\ell'$ and since $j_{\ell'}(F)\propto F^{\ell'}$ for
small $F$.   
Likewise, Faraday rotation induces E-B correlations  with $\ell_2 - \ell_1$ 
odd. It also distorts the diagonal $\Theta$-E and E-E correlations, 
and induces non-diagonal terms with $|\ell_2-\ell_1|=2k$, proportional at 
least to $F^{2k}$ for small $F$. 

In the system with $\hat e_3\parallel \vec B$, 
we can actually calculate to all orders the effect of Faraday rotation
upon the diagonal 
part of the correlation $\Theta$-E. Indeed, since  $\sum_{m}
R(\ell m,\ell',\ell m)=(2\ell+1)\delta_{\ell '0}$ we find that 
\begin{equation}
C_\ell^{\Theta E}\equiv \frac{1}{2\ell +1}\sum_{m=-\ell}^{\ell}<\Theta^*_{\ell m} 
E_{\ell m}>= C_\ell^{\Theta \tilde E^{(0)}}\, .
\label{TE}
\end{equation}

\begin{figure}[t]
\includegraphics[width=3.in,angle=-90]{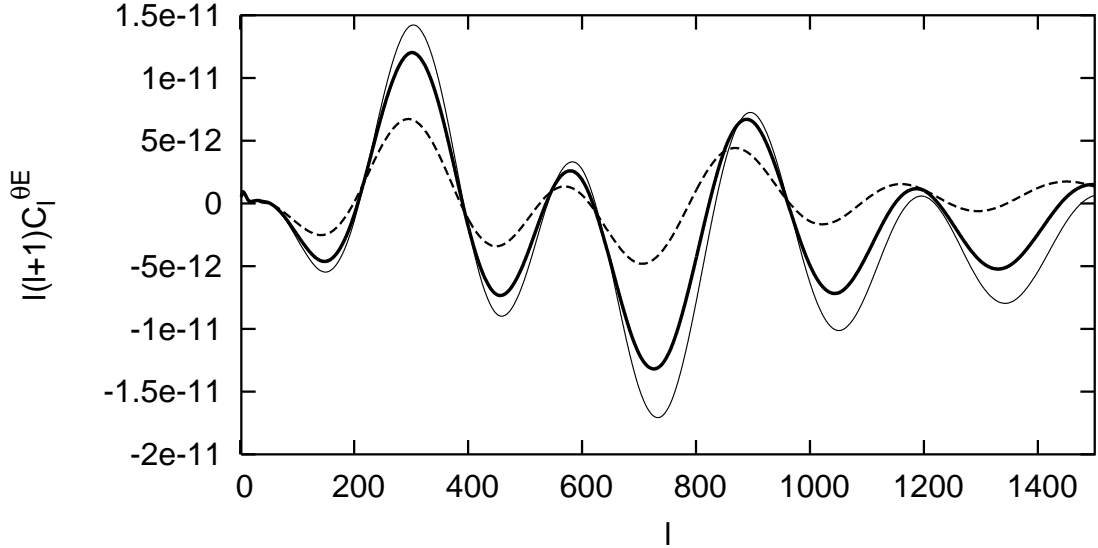}
\caption{Temperature-polarization (E-mode) cross-correlation in a $\Lambda$CDM 
background cosmology in the case of no magnetic field (thin solid line) and in 
the case of Faraday rotation in a uniform magnetic 
field with $F=1$ (thick solid line) and $F=3$ (dashed line). $F$ is defined in 
eq.~(\ref{F}). 
For large $F$ the depolarizing effect of differential Faraday rotation is 
quite noticeable.}
\label{fig.clte} 
\end{figure}

In other words, the diagonal terms in the cross-correlation $\Theta$-E 
can be calculated, to all orders,
exactly as in the case of no magnetic field with the addition of a term 
$j_0(F\tau)$ in the integrand
along the line of sight for $E_{\ell m}$. Notice however that this equivalence 
is not valid for individual contributions to $C_\ell$ with different values of $m$,
since they also get additional contributions from higher $\tilde E_{\ell 
m}^{(\ell ')}$ which, however, add to zero in the sum over $m$. 
We can readily evaluate expression (\ref{TE}) numerically 
using CMBFAST \cite{se96}, account taken of the extra term $j_0(F\tau)$. 
As background cosmology a spatially-flat $\Lambda$-CDM model was assumed, with 
70\% of the energy 
density in the form of a cosmological constant, 25.6\% in cold dark matter, 
4.4\% in baryons, a Hubble 
constant $H_0=71$~km/sec Mpc, adiabatic scale invariant scalar fluctuations, 
and an optical depth since 
reionization $\tau_{\rm reion}=0.17$, as suggested by WMAP measurements of 
temperature-polarization 
correlations \cite{kogut03}. The result is plotted in Figure~\ref{fig.clte}, in 
the standard case of no 
Faraday rotation (thin solid  line), and in the  case of Faraday rotation 
factors $F=1$ (thick solid line) and $F=3$ (dashed line).

\begin{figure}[t]
\includegraphics[width=4.in,angle=-90]{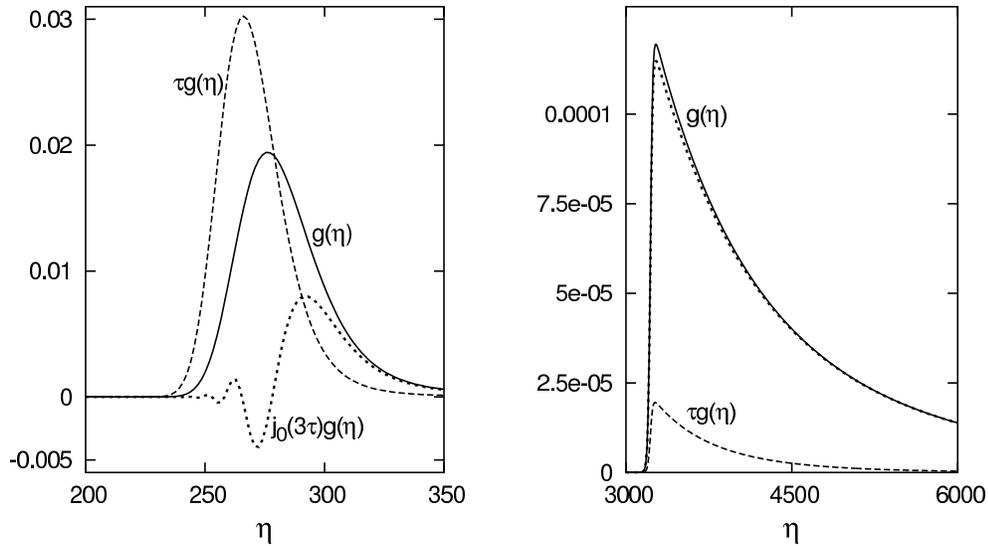}
\caption{Visibility function $g(\eta)$ (solid line) as a function of 
conformal time in the background $\Lambda$CDM cosmology, calculated with 
CMBFAST, and multiplied by $j_0(F\tau)$ with $F=3$ (dotted line), and 
multiplied by $\tau$ (dashed line). The left panel displays times around 
hydrogen recombination and photon decoupling. The right panel displays 
times after reionization, which was assumed to  take place at an optical 
depth $\tau_{\rm reion}=0.17$.}
\label{fig.visi} 
\end{figure}

Notice that there is no linear effect of Faraday rotation upon 
$C_\ell^{\Theta E}$ for small $F$, 
the lowest order correction being quadratic and with negative sign. 
For large $F$ the oscillatory nature of $j_0(F\tau)$ leads to a significant
reduction in the amplitude of the cross-correlation, reflecting the depolarizing effect of 
large differential Faraday rotation \cite{ha97}. A shift towards larger angular 
scales (smaller $\ell$) in the positions of the peaks is also 
noticeable in Fig.~\ref{fig.clte}. 
This can be interpreted as the consequence of the 
fact that the integrand in eq.~(\ref{Etildeprime}) for $\tilde E_{\ell m}^{(0)}$ 
contains the factor $g(\eta)j_0(F\tau)$, where
$g(\eta)=\dot\tau\exp{(-\tau)}$ is the visibility function (the probability that a 
photon last-scattered 
within ${\rm d}\eta$ of $\eta$). The visibility function around 
the time of hydrogen recombination can be well approximated by a Gaussian centered at
the time of photon decoupling. The polarization produced by Thomson scattering at recombination
can be well approximated as proportional to the time derivative of temperature inhomogeneities around decoupling 
and to the width of the visibility function. The additional factor $j_0(F\tau)$ in the integrand of 
eq.~(\ref{Etildeprime}) not only reduces the value of
the integral, but also shifts to later times the location of the largest contribution. 
We display this effect in Fig.~\ref{fig.visi}, where we plot the visibility function $g(\tau)$ (solid line) 
and the factor $g(\eta)j_0(F\tau)$ for $F=3$ (dotted line), both around the time of photon decoupling 
(left panel) and  after reionization (right panel). The shift to later times in the location of the peak 
in $g(\eta)j_0(F\tau)$ in the left panel translates into a shift towards larger angular scales in the 
position of the peaks in the correlation function.

Faraday rotation has a relatively smaller effect upon $E$ modes produced after reionization
because $\tau$ is not very large (smaller than 0.17 in our
numerical examples), and thus $j_0(F\tau)$ is closer to unity unless $F$ is larger.

The strong damping of $\Theta$E correlations that occurs when $F$ is of order unity or larger 
also affects
$\Theta$B, EB, EE and BB correlations. It is a consequence that differential 
Faraday rotation across the last scattering surface (or after reionization if $F$ were
large enough) has a net depolarizing effect.

Corrections to $C_\ell^{\Theta E}$  smaller than those considered 
here exist due to the fact that Faraday rotation affects the quadrupole term in the 
temperature anisotropy source \cite{ha97}, an effect that we have neglected. 

It is clear from Fig.~\ref{fig.clte} that Faraday rotation 
factors of order unity would imprint
a measurable signature upon the temperature-polarization cross-correlation, which is 
moreover frequency-dependent. Primordial magnetic fields with strength $10^{-8}$ G 
(scaled to present values) 
can easily be ruled out with measurements at frequencies of order 30 GHz. 

\section{Correlation functions for small Faraday rotation}

In the (likely to be realistic) case of small Faraday rotation ($F<<1$), 
it suffices to take $\ell'=1$ in eq. (\ref{cross}) or
(\ref{crossrot}). To this order, the cross-correlation $\Theta$-B in
the system with $\hat e_3\parallel \vec B$ reads
\begin{equation}
<\Theta^*_{\ell_1 m}B_{\ell_2 m}>=C_{\ell_1}^{\Theta \tilde E^{(1)}}
\left[\delta_{\ell_2,\ell_1+1}\frac{\sqrt{(\ell_1+1)^2-m^2}\sqrt{(\ell_1+1)^2-4}}
{(\ell_1+1)\sqrt{2\ell_1+1}\sqrt{2\ell_1+3}}+
\delta_{\ell_2,\ell_1-1}\frac{\sqrt{\ell_1^2-m^2}\sqrt{\ell_1^2-4}}
{\ell_1\sqrt{2\ell_1-1}\sqrt{2\ell_1+1}}\right]\, .
\label{TB}
\end{equation}
In a general frame there are also correlations among multipoles with
values of $m$ differing by $\pm 1$. 
Notice that a similar expression holds for the cross correlation E-B, with
$C_{\ell_1}^{\tilde E^{(0)} \tilde E^{(1)}}$ in place of
$C_{\ell_1}^{\Theta \tilde E^{(1)}}$.

\begin{figure}[t]
\includegraphics[width=3.in,angle=-90]{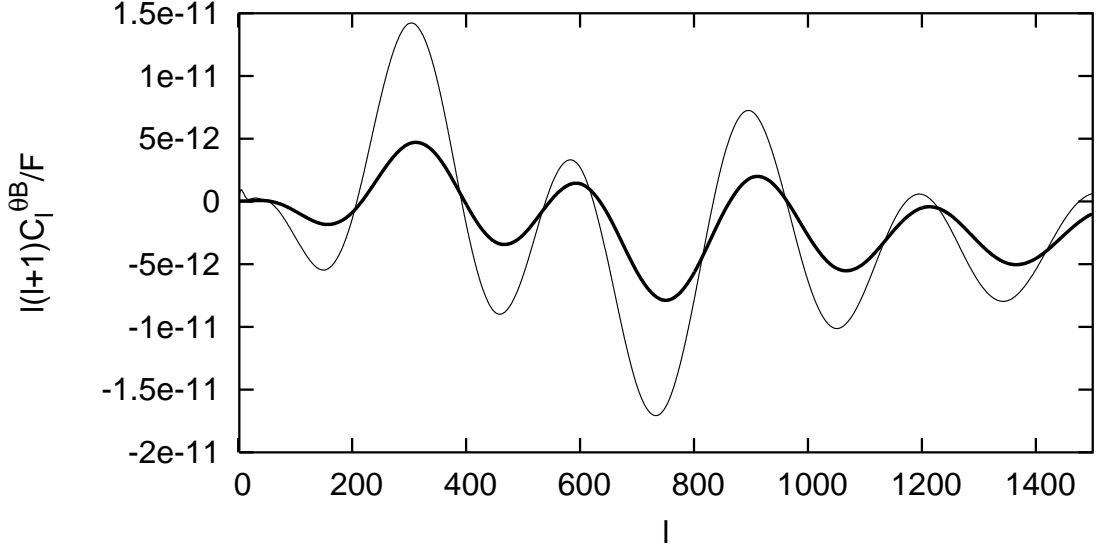}
\caption{Temperature-polarization (B-mode) cross-correlation (thick solid line) in 
a $\Lambda$CDM background cosmology due to Faraday rotation of E-modes, scaled by the 
factor $F$ defined in 
eq.~(\ref{F}), assumed to be smaller than unity. See the text for the definition of 
$C_l^{\Theta{\rm B}}$, which is actually a non-diagonal correlation. The correlation 
$\Theta$-E in the absence of magnetic field is also displayed for reference 
(thin solid line).}
\label{fig.cltb} 
\end{figure}

In Fig.~\ref{fig.cltb} we plot, as a measure of the strength of the cross-correlations 
$\Theta$-B, the quantity 
\begin{equation}
C_{\ell}^{\Theta B}\equiv C_{\ell}^{\Theta \tilde E^{(1)}}
\frac{1}{2\ell +1}\sum_{m=-\ell}^{\ell}\frac{\sqrt{\ell^2-m^2}\sqrt{\ell^2-4}}{
\ell\sqrt{2\ell-1}\sqrt{2\ell+1}}\, ,
\end{equation}
which represents an average over $m$ of the
amplitude of the non-diagonal correlations. We plot it scaled down by the factor $F$,
assumed to be small.
Notice that $C_{\ell}^{\Theta \tilde E^{(1)}}$ coincides with the power spectrum for 
$\Theta$-E correlations produced by scalar fluctuations 
in the absence of a magnetic field except for an extra factor $3 j_1(F\tau)\approx 
F\tau$ in the time integral that leads to E. This can be interpreted 
as the fact that the B-modes are the result of the E-modes produced by Thomson scattering up to a time $\eta$, 
further Faraday rotated by a factor $F\tau$ since, and integrated over $\eta$. Fig.~\ref{fig.visi} displays 
the factors $g(\eta)$ and  $\tau g(\eta)$ in the background cosmology chosen for illustration, both around photon 
decoupling (left panel) and after reionization (rigth panel). $\tau g(\eta)$ has its peak shifted to earlier times 
around decoupling compared with the visibility function. This effect shifts the peaks in the correlation functions 
to smaller angular scales (larger values of $\ell$). Faraday rotation effects after reionization are not only 
suppressed by a factor $F$, but also by the relatively smaller values of $\tau$, as displayed in the right panel.

The diagonal contribution to the angular autocorrelation of B-modes is given, to 
lowest order in $F$ and averaged over $m$, by  

\begin{equation}
C_\ell^{BB}\equiv\frac{1}{2\ell + 1}\sum_{m=-\ell}^{\ell}<B^*_{\ell m}B_{\ell 
m}>=\frac{1}{3}\frac{(\ell+1)^2-4}{(2\ell +3)(\ell+1)}
C_{\ell+1}^{\tilde E^{(1)}\tilde E^{(1)}}+
\frac{1}{3}\frac{\ell^2-4}{(2\ell +1)\ell}
C_{\ell-1}^{\tilde E^{(1)}\tilde E^{(1)}}\, .
\end{equation}

\begin{figure}[t]
\includegraphics[width=4.in,angle=-90]{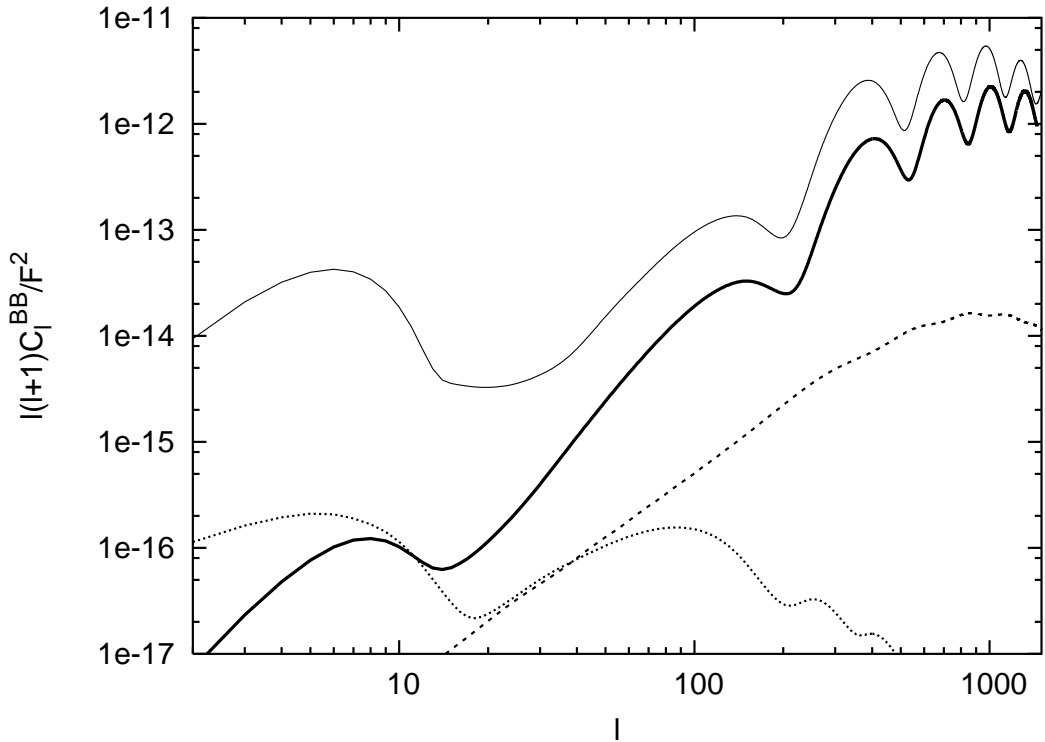}
\caption{Angular autocorrelation multipoles of B-modes of CMB polarization generated by Faraday rotation of E-modes 
generated by scalar perturbations in a $\Lambda$CDM cosmology, scaling away the factor $F^2$ (thick solid line). The 
autocorrelation of E-modes from which they originate is also displayed (thin solid line) for comparison purposes. Also 
shown are the autocorrelation of B-modes due to weak gravitational lensing (dashed line) and to gravitational waves in 
a model with tensor to scalar ratio $r=10^{-2}$ (dotted line).}
\label{fig.clbb} 
\end{figure}

We calculate it numerically using CMBFAST \cite{se96}, accounting for the extra 
factors $F\tau$ in the integrals along 
the line of sight for $\tilde E^{(1)}$. The result is displayed 
in Figure~\ref{fig.clbb}, for the same background 
$\Lambda$-CDM cosmological model as in the
previous section. The quantity plotted (thick solid line) is 
$\ell(\ell+1)C_\ell^{BB}/F^2$, with $F$ given by eq.~(\ref{F}), so
that the result should be scaled down by the appropriate value of
$F^2$, which we have assumed to be small. We also plot, for comparison purposes, the B-mode correlation due to weak 
gravitational lensing (dashed line), and that due to gravitational waves in a model with tensor to scalar ratio 
$r=10^{-2}$ (following the convention in e.g. \cite{peiris03} for its definition). Also plotted in Fig.~\ref{fig.clbb}, 
to help estimate the strength of the effect, is $\ell(\ell+1)C_\ell^{EE}$ due to the scalar fluctuations in the absence 
of Faraday rotation (thin solid line).

The angular correlation multipoles of B-modes are roughly a factor $F^2/3$ smaller than those
of the E-modes from which they originate by Faraday rotation, at relatively large $\ell$.
Notice also the shift in the peaks positions towards larger $\ell$.
At small $\ell$ the effect is relatively suppressed because  the B-modes produced by Faraday rotation are smaller than 
the E-modes generated after reionization by an extra factor somewhat smaller than $\tau_{\rm reion}^2$ 
(the optical depth to reionization).

It is clear from Fig.~\ref{fig.clbb} that Faraday rotation factors of 
order $F=0.1$, which
at $\nu=$~30~GHz correspond to a primordial magnetic field of present 
strength $B_0=10^{-9}$~G,
induce B-modes with
amplitude comparable to those generated by weak gravitational lensing.
B-modes due to Faraday rotation in the magnetic fields of clusters of 
galaxies could have a
comparable strength (but different angular spectrum) around $\ell=1000$ 
only if $B_{\rm clusters}\approx 2~\mu$G,
uniform across the clusters \cite{ta01,oh03}.

\section{Conclusions}

The detection of polarization in the CMB \cite{dasi} and its
correlation to the temperature anisotropies \cite{kogut03} has become
a very promising tool to test the early universe. Much observational
effort is taking place to measure $\Theta$, E and B correlation
functions. The cross-correlation function $\Theta$E already gives
information on the reionization history of the universe. The 
B-mode polarization is being looked upon specially as a way to probe
a primordial gravitational wave background which would represent a clear
signature of a period of inflation in the early Universe.
Precision data on the CMB polarization is expected from the Planck
Mission \cite{planck} and future dedicated missions, such as NASA's 
{\it Beyond Einstein} Inflation Probe \cite{IP} or ground-based 
experiments, like {\it BICEP} \cite{BICEP} and {\it PolarBeaR}.

There are other processes beside a gravitational wave background that
can originate B-mode polarization, more prominently the gravitational 
lensing conversion of a fraction of the dominant E-type  into B-type
polarization, which can however be significantly cleaned from the data 
allowing to probe inflationary models with tensor-to-scalar ratios $r
\le 10^{-6}$\cite{seljak03}.

We have studied here in detail the effect of the Faraday rotation
produced by a large scale uniform primordial magnetic field upon the
polarization of the CMB for the case of scalar density
perturbations. 
The exact solution for the E and B modes of polarization in the presence
of a uniform magnetic field is given in eq.~(\ref{EBfinal}).
Faraday rotation of the direction of the polarization
generates a B-polarization component from the initial E-modes, which correlates
with temperature anisotropies and E-polarization \cite{sc97}. 
$\Theta$-B correlations are non-diagonal, and their strength is of
order $F$ times the original $\Theta$-E correlations, with 
the parameter $F$ defined by eq.~(\ref{F}), as long as $F$ is smaller 
than unity. 
The amplitude of the B-polarization angular autocorrelation function
generated by Faraday rotation is comparable
to the (uncleaned) weak lensing signal if $F\sim0.1$,
which corresponds to a magnetic field $B\simeq 10^{-9}$G at a
frequency $\nu \simeq 30$GHz. Let us notice however, that the lensing
signal, as well as that generated by gravitational waves, are independent 
of the frequency, while F scales as $\nu^{-2}$, thus multi-frequency 
measurements should easily distinguish the Faraday rotation contribution.

Small Faraday rotation factors have a relatively small effect upon $\Theta$E
and EE correlations, since they induce distortions roughly $F^2$ smaller than
their values in the absence of a magnetic field. If $F$ were of order unity
or larger, all correlations $\Theta$E, $\Theta$B, EE, EB and BB
are significantly damped, due to the additional 
depolarizing effect of
differential Faraday rotation across the last scattering surface. The good
agreement between WMAP measurements \cite{kogut03} of $\Theta$E correlations 
and its predicted value in a standard cosmology rules out values of $F$ of order 
unity, which implies primordial magnetic fields of present strength 
$B_0\approx 10^{-8}$G at frequencies $\nu=$30 GHz. Future precision measurements 
should allow to probe smaller values of primordial magnetic fields through 
Faraday rotation effects.

\appendix*
\section{Boltzmann equations for E and B}

We write down here the Boltzmann equations for the E and B-modes of polarization including 
the Faraday rotation effect of a constant magnetic field. We adopt the formalism of Ref. \cite{hw}
for the case without a magnetic field. The additional terms, that mix E and B modes, can be obtained 
by derivation of the defining equation (\ref{EB}), using eq. (\ref{QU}), and the 
Clebsh-Gordan relations for products of spherical harmonics.     
It turns out that

\begin{eqnarray}
\dot E_{\ell}^{(m)} &=& 
  k \Bigg[ 
        \frac{\kap{2}{\ell}{m}} {(2\ell-1)}
E_{\ell-1}^{(m)} - \frac{2m}{\ell (\ell + 1)} B_\ell^{(m)} 
- \frac{\kap{2}{\ell+1}{m}}{ (2 \ell + 3)}  E_{\ell + 1}^{(m)} \Bigg]- \dot\tau [E_\ell^{(m)} + 
\sqrt{6} P^{(m)} \delta_{\ell,2}] \nonumber\\
& & + iF \dot\tau
\left[ \cos \theta_B \left(-\frac{\kap{2}{\ell+1}{m}}{2\ell+3} \; B_{\ell +1}^{(m)} +2\frac{m}
{\ell(\ell+1)}\;E_\ell^{(m)}+ \frac{\kap{2}{\ell}{m}}{2\ell-1}\;B_{\ell -1}^{(m)} \right) \right. \nonumber\\
&&\left. -\frac{1}{\sqrt{2}}\sin{\theta_B}e^{-i \phi_B} \left(\frac{\gama{2}{\ell+1}{-m}}{2\ell+3}\; 
B_{\ell +1}^{(m-1)}-2\alfa{\ell}{m-1}\; E_{\ell}^{(m-1)}+\frac{\gama{2}{\ell}{m-1}}{2\ell-1}\; 
B_{\ell -1}^{(m-1)}\right) \right.\nonumber\\
&&\left.+\frac{1}{\sqrt{2}}\sin{\theta_B}e^{i \phi_B} \left(\frac{\gama{2}{\ell+1}{m}}{2\ell+3}\; 
B_{\ell+1}^{(m+1)}+2 \alfa{\ell}{m}\; E_{\ell}^{(m+1)}+\frac{\gama{2}{\ell}{-(m+1)}}{2\ell-1}\; 
B_{\ell-1}^{(m+1)} \right) \right]
\end{eqnarray} 
and 
\begin{eqnarray}
\dot B_{\ell}^{(m)} &=& k \Bigg[ \frac{\kap{2}{\ell}{m}}{ (2\ell-1)}
  B_{\ell-1}^{(m)} + \frac{2m}{\ell (\ell + 1)} E_\ell^{(m)} -
  \frac{\kap{2}{\ell+1}{m}}{ (2 \ell + 3)} B_{\ell + 1}^{(m)} \Bigg]-
  \dot\tau B_\ell^{(m)} \nonumber\\
 & & + iF  \dot\tau 
\left[ \cos \theta_B \left(\frac{\kap{2}{\ell+1}{m}}{2\ell+3} \; E_{\ell +1}^{(m)} +2\frac{m}{\ell(\ell+1)}\;
B_\ell^{(m)}- \frac{\kap{2}{\ell}{m}}{2\ell-1}\;E_{\ell -1}^{(m)} \right) \right. \nonumber\\
&&\left. -\frac{1}{\sqrt{2}}\sin{\theta_B}e^{-i \phi_B} \left(-\frac{\gama{2}{\ell+1}{-m}}{2\ell+3}\; 
E_{\ell +1}^{(m-1)}-2\alfa{\ell}{m-1}\; B_{\ell}^{(m-1)}-\frac{\gama{2}{\ell}{m-1}}{2\ell-1}\; 
E_{\ell -1}^{(m-1)}\right) \right.\nonumber\\
&&\left.+\frac{1}{\sqrt{2}}\sin{\theta_B}e^{i \phi_B} \left(-\frac{\gama{2}{\ell+1}{m}}{2\ell+3}\; 
E_{\ell+1}^{(m+1)}+2 \alfa{\ell}{m}\; B_{\ell}^{(m+1)}-\frac{\gama{2}{\ell}{-(m+1)}}{2\ell-1}\; 
E_{\ell-1}^{(m+1)} \right) \right]
\end{eqnarray} 
where $\theta_B$ and $\phi_B$ are the polar and azimuthal angles respectively
of the direction of the
magnetic field in a system with $\hat e_3\parallel\vec k$, 
and we have defined  

\begin{eqnarray}
\kap{s}{\ell}{m} &=& \frac{\sqrt{(\ell^2-m^2)(\ell^2-s^2)}}{\ell} \nonumber\\
\gama{s}{\ell}{m} &=& \frac{\sqrt{(\ell^2-s^2)(\ell+m)(\ell+m+1)}}{\sqrt{2}\ell} \nonumber\\
\alfa{\ell}{m} &=& \frac{\sqrt{(\ell-m)(\ell+m+1)}}{\sqrt{2}\ell(\ell+1)} \nonumber
\end{eqnarray}

\acknowledgments{We thank Anthony Challinor for pointing us a mistake in a 
previous version of this paper. Work partially supported by ANPCyT, Fundaci\'on Antorchas and CONICET.}


\begin{thebibliography}{99}

\bibitem{se97}
U.~Seljak and M.~Zaldarriaga,
Phys.\ Rev.\ Lett.\  {\bf 78}, 2054 (1997); 
ibid, Phys.\ Rev.\ D {\bf 55}, 1830 (1997).

\bibitem{ka97}
M.~Kamionkowski, A.~Kosowsky and A.~Stebbins,
Phys.\ Rev.\ Lett.\  {\bf 78}, 2058 (1997);
ibid, Phys.\ Rev.\ D {\bf 55}, 7368 (1997). 

\bibitem{zalsel98}
M.~Zaldarriaga and U.~Seljak,
Phys.\ Rev.\ D {\bf 58}, 023003 (1998).

\bibitem{knox}
L.~Knox and Y.~S.~Song,
{\it et al.},
Phys.\ Rev.\ Lett.\  {\bf 89}, 011303 (2002).

\bibitem{kesden}
M.~Kesden, A.~Cooray and M.~Kamionkowski,
Phys.\ Rev.\ Lett.\  {\bf 89}, 011304 (2002); ibid,
Phys.\ Rev.\ D {\bf 67}, 123507 (2003).

\bibitem{hirata}
C.~M.~Hirata and U.~Seljak,
Phys. Rev. D {\bf 68}, 083002 (2003).

\bibitem{seljak03}
U.~Seljak and C.~M.~Hirata,
Phys. Rev. D {\bf 69}, 043005 (2004).

\bibitem{hu00}
W.~Hu
Astrophys.\ J.\  {\bf 529}, 12 (2000).

\bibitem{mo04}
S.~Mollerach, D.~Harari and  S.~Matarrese, 
Phys. Rev. D {\bf 69}, 063002 (2004).

\bibitem{se01}
T.~R.~Seshadri and K.~Subramanian,
Phys.\ Rev.\ Lett.\  {\bf 87}, 101301 (2001).

\bibitem{ma02}A.~Mack, T.~Kahniashvili, A.~Kosowsky,
Phys. Rev. D {\bf 65}, 123004 (2002).

\bibitem{su03}
K.~Subramanian, T.~R.~Seshadri and J.~D.~Barrow,
Mon.\ Not.\ Roy.\ Astron.\ Soc.\ {\bf 344}, L31 (2003).

\bibitem{ko96}
A.~Kosowsky and A.~Loeb,
Astrophys. J. {\bf 469}, 1 (1996).

\bibitem{sc97}E. S. Scannapieco and P. G. Ferreira,
Phys. Rev. D {\bf 56}, R7493 (1997).

\bibitem{po02}L. Pogosian, T. Vachaspati and S. Winitzki, Phys. Rev. D 
{\bf 65} 083502 (2002).

\bibitem{ca03}C.~Caprini, R.~Durrer and T.~Kahniashvili,
Phys. Rev. D {\bf 69} 063006 (2004).

\bibitem{lu99}A.~Lue, L.~Wang and M.~Kamionkowski, Phys. Rev. Lett. 
{\bf 83}, 1506 (1999); K.~R.~S.~Balaji, R.~H.~Brandenberger and 
D.~A.~Easson, JCAP {\bf 12}, 008 (2003).

\bibitem{kr94}
P.~P.~Kronberg, Rep. Prog. Phys. {\bf 57}, 325 (1994).

\bibitem{wi02}
L. M. Widrow,
Rev. of Mod. Phys. {\bf 74}, 775 (2002).

\bibitem{dyn}
E.~N.~Parker, Astrophys. J. {\bf 163}, 255 (1971), A.~A.~Ruzmaikin,
A.~M.~Shukorov and D.~D.~Sokoloff, {\it Magnetic Fields in Galaxies},
(Kluwer, Dordretch, 1988).
 
\bibitem{prim}
F.~Hoyle, {\it La structure et l'evolution de l'universe}, XI Solvay
conference (Brussels: R.~Stoops), 59; J.~H.~Piddington, Mon. Not. of
R. Astron. Soc. {\bf 128}, 345 (1964), A.~M.~Howard and R.~M.~Kulsrud, 
Astrophys.\ J.\  {\bf 483}, 648 (1997).

\bibitem{ta01}M.~Takada, H.~Ohno and N.~Sugiyama, ``Faraday rotation 
effect of intracluster magnetic field on cosmic microwave background 
polarization'', astro-ph/0112412.

\bibitem{oh03}H.~Ohno, M.~Takada, K.~Dolag, M.~Bartelmann and N.~Sugiyama, 
Astrophys. J. {\bf 584}, 599 (2003).

\bibitem{chandra} S. Chandrasekar, {\it Radiative Transfer} 
(Dover, New York, 1960).

\bibitem{ha97}
D.~Harari, J.~D.~Hayward and M.~Zaldarriaga,
Phys.\ Rev.\ D {\bf 55}, 1841 (1997). 

\bibitem{zh95}
M.~Zaldarriaga and D.~D.~Harari,
Phys.\ Rev.\ D {\bf 52}, 3276 (1995).

\bibitem{hw}
W.~Hu and M.~White,
Phys.\ Rev.\ D {\bf 56}, 596 (1997). 

\bibitem{hu98}
W.~Hu, U.~Seljak, M.~White and M.~Zaldarriaga,
Phys.\ Rev.\ D {\bf 57} 3290 (1998). 

\bibitem{se96}
U.~Seljak and M.~Zaldarriaga,
Astrophys.\ J.\  {\bf 469}, 437 (1996).

\bibitem{dasi}J. Kovac {\it et al.}, Nature {\bf 420}, 772 (2002).

\bibitem{kogut03}A. Kogut {\it et al.},
Astrophys.\ J.\ Suppl.\  {\bf 148}, 161 (2003).

\bibitem{peiris03}
H.~V.~Peiris {\it et al.},
Astrophys.\ J.\ Suppl.\  {\bf 148}, 213 (2003).

\bibitem{planck}J. Tauber, Advances in Space Research {\bf 34}, 
491 (2004);see also http://astro.estec.esa.nl/Planck.

\bibitem{IP}See http://universe.gsfc.nasa.gov

\bibitem{BICEP}B.~G.~Keating et. al, in "Polarimetry in Astronomy", 
edited by S. Fineschi, Proceedings of the SPIE, Volume 4843, pp. 284 
(2003); see also http://cosmology.berkeley.edu/group/swlh/bicep.

\end{thebibliography}
\end{document}